\documentstyle[prl,aps,epsf,multicol]{revtex}
\begin{document}
\title{\bf Survival of a Diffusing Particle in a Transverse Flow Field}
\author{Alan J. Bray and Panos Gonos}
\address{School of Physics and Astronomy, University of 
Manchester, Manchester M13 9PL, UK}

\date{\today}

\maketitle

\begin{abstract}
We consider a particle diffusing in the $y$-direction, $dy/dt=\eta(t)$
where $\eta(t)$ is  Gaussian white noise, and subject  to a transverse
flow field  in the  $x$-direction, $dx/dt=f(y)$, where  $x \ge  0$ and
$x=0$ is an absorbing boundary.  We discuss the time-dependence of the
survival probability of  the particle for a class  of functions $f(y)$
that are positive in some regions of space and negative in others.
\noindent

\end{abstract}

\begin{multicols}{2}

\section{Introduction}

In  a recent  paper \cite{GB04}  we considered  a class  of stochastic
processes defined by the equations
\begin{eqnarray}
\label{langevin1}
\dot{y} & = & \eta(t)\ , \\ \dot{x} & = & f(y),
\label{langevin2}
\end{eqnarray}
where $\eta(t)$ is Gaussian white  noise with mean zero and correlator
$\langle    \eta(t)\eta(t')\rangle   =    2D\delta(t-t')$.   Equations
(\ref{langevin1})  and (\ref{langevin2})  represent  a particle  which
diffuses (performs a random walk)  in the $y$-direction but is subject
to a deterministic drift in the $x$-direction with a $y$-dependent drift
velocity. We consider the case where there is an absorbing boundary at
$x=0$,  and we  are  interested  in the  survival  probability of  the
particle.  Two  previously studied models  are the cases  $f(y)=y$ and
$f(y)={\rm  sgn}(y)$.   The  former   is  equivalent  to   the  random
acceleration  process,  $\ddot{x}=\eta(t)$, while  the  latter is  the
`windy cliff' model introduced  by Redner and Krapivsky \cite{RK}. For
both these models, the survival  probability of the particle decays as
$t^{-1/4}$ for large  time $t$, and it has  been argued \cite{GB04,RK}
that this is a generic result  for odd functions $f(y)$. It should
be noted  that, in  general, such problems  are highly  nontrivial and
exact solutions are available in only a small number of cases.

In ref.\cite{GB04},  we raised  the question of how the survival
probability decays  when the function  $f(y)$ is not an  odd function.
We considered a  class of models where $f(y)=v_+  y^\alpha$ for $y>0$,
and  $f(y)= -v_-(-y)^\alpha$  for  $y<0$.  Here  the  drift takes  the
particle away  from the absorbing  boundary for $y>0$, and  towards it
for $y<0$,  but not in a  symmetrical way: the function  $f(y)$ is odd
only when  $v_+=v_-$. Using an  extension of a technique  developed by
Burkhardt \cite{Burkhardt} in  connection with the random acceleration
process,   we  showed   that  the  survival   probability  decays   as
$t^{-\theta}$  for  large  $t$,   where  the  exponent  $\theta$  (the
`persistence exponent') is nontrivial and given by
\begin{equation}
\theta = \frac{1}{4}-\frac{1}{2\pi\beta}\tan^{-1}
\left[\frac{v_+^\beta - v_-^\beta}{v_+^\beta + v_-^\beta}
\tan\left(\frac{\pi\beta}{2}\right)\right]
\label{result}
\end{equation}
where  $\beta  =  1/(2+\alpha)$.   This  reduces  to  $\theta=1/4$  for
$v_+=v_-$   but  $\theta$   lies  in   the  range   $0<\theta<1/2$  in
general.  When $v_+>v_-$,  the drift  away from  the boundary  is {\em
weakly} dominant.  The decay of  the survival probability still  has a
power law form, but $\theta < 1/4$, while the converse occurs for $v_+
<  v_-$:  the  drift  toward  the  boundary  is  weakly  dominant  and
$\theta>1/4$. The  fact that $1/2$ is  an upper bound  for $\theta$ is
clear when one recalls that the probability that the particle stays in
the region $y>0$ (and therefore never encounters the drift towards the
absorbing boundary) decays as $t^{-1/2}$ \cite{Guide,majumdar_review}.

It   is  worth  emphasising   that  the   process  defined   by  Eqs.\
(\ref{langevin1})  and (\ref{langevin2})  is non-Gaussian  except when
$f(y)$ is linear.  Eq.\ (\ref{result}) is a rare example of an exactly
calculable persistence exponent for a non-Gaussian process.
  
In \cite{GB04}  we speculated about  the behavior of the  system when
the  function  $f(y)$  is  described  by  different  {\em  exponents},
$\alpha_+$ and $\alpha_-$, for $y>0$ and $y<0$ respectively. We argued
that for $\alpha_+>\alpha_-$ the drift  away from the boundary is {\em
strongly}  dominant, leading  to  a non-zero  survival probability  at
infinite time (i.e.\ $\theta=0$) , while for $\alpha_+ < \alpha_-$ the
drift towards  the boundary is strongly dominant  and $\theta$ obtains
its maximum value of $1/2$.

In  the present  work we  obtain some  exact result  pertinent  to the
latter question.  Looking  at the case $\alpha_+ >  \alpha_-$, we show
explicitly that  the survival probability approaches  a non-zero value
at  infinite time  and we  obtain  a closed-form  expression for  this
value.

In the  second part of the paper  we revisit the issue  of whether the
decay  exponent $\theta$  takes  the  value $1/4$  for  {\em all}  odd
functions $f(y)$, as  has been assumed up to  now.  We consider models
where the flow field takes a periodic, banded form  with $f(y)$ taking
the values $v_+$  and $-v_-$ in alternate bands  of width $b_+$, $b_-$
respectively. We  demonstrate numerically that for  the case $b_+=b_-$
and  $v_+ = v_-$  which, with  appropriate cloice  of the  $y=0$ axis,
represents an  odd function,  $\theta = 1/2$,  not $1/4$ and  we argue
that $\theta=1/2$ whenever there is no net drift, i.e.\ when $v_+b_+ =
v_-b_-$. In  these cases we argue  that an effective  diffusion in the
$x$-direction,  resulting  from the  stochastic  movement between  the
alternating bands, dominates over the drift and naturally accounts for
the exponent $1/2$.  We obtain an analytical result  in the limit $b_-
\to  0$, $v_-  \to  \infty$, with  $b_-v_-$  held fixed  and equal  to
$b_+v_+$, and we verify that $\theta=1/2$ in this case.

   
\section{Model and Calculation}

The model we consider initially is defined by Eqs.\ (\ref{langevin1}) 
and (\ref{langevin2}), with the function $f(y)$ given by
\begin{equation}
f(y) = \cases{v_+ y^{\alpha_+},\ \ \ \ \ \ \ \ \ \ y>0, \cr
-v_-(-y)^{\alpha_-},\ \ \ y<0.\cr}
\label{f}
\end{equation}
We  work  in the  regime  $\alpha_+ >  \alpha_-$,  where  we expect  a
non-zero  survival  probability. The  probability,  $Q(x,y)$ that  the
particle   survives   to   infinite   time  satisfies   the   backward
Fokker-Planck  equation (BFPE) $D\partial_{yy} Q  + f(y)\partial_xQ=0$.  
In the present context it is  convenient to work instead with the `killing
probability', $P=1-Q$, which obviously satisfies the same equation,
\begin{equation}
D \partial_{yy}P + f(y)\,\partial_xP = 0\ ,
\label{BFPE}
\end{equation}
but with different boundary conditions. The boundary conditions on 
$P(x,y)$ are 
\begin{eqnarray}
\label{bc1}
P(\infty, y) & = & 0, \\
\label{bc2}
P(x,\infty) &=& 0, \\
\label{bc3}
P(0,y) & = & 1,\ \ y<0\ ,
\end{eqnarray}
the  last  of these  following  from the  fact  that,  for an  initial
condition  with  $x=0$  and  $y<0$,  the flow  immediately  takes  the
particle on to the absorbing boundary.

Inserting the form (\ref{f}) for $f(y)$ in (\ref{BFPE}), the equation can 
be solved separately in each regime by separation of variables. Imposing 
the boundary conditions (\ref{bc1}) and (\ref{bc2}) the result 
takes the form (setting $D=1$ for convenience)
\begin{eqnarray}
\label{possoln}
P(x,y) &=& \sqrt{y}\int_0^\infty dk\,a(k)K_{\beta_+}
\left(2\beta_+\sqrt{kv_+}y^{1/2\beta_+}\right)\,e^{-kx}, \nonumber \\
&&\hspace{4.5cm} {\rm for}\ y>0, \\
&=& \sqrt{-y}\int_0^\infty dk\,\left\{b(k)J_{\beta_-}
\left(2\beta_-\sqrt{kv_-}(-y)^{1/2\beta_-}\right)\right. \nonumber \\
&& \hspace{0.2cm} + \left. c(k)J_{-\beta_-}\left(2\beta_-\sqrt{kv_-}
(-y)^{1/2\beta_-}\right)\right\}\,e^{-kx}, \nonumber \\
&&\hspace{4.5cm} {\rm for}\ y<0, 
\label{negsoln}
\end{eqnarray}
where 
\begin{equation}
\beta_\pm = 1/(2+\alpha_\pm), 
\end{equation}
$J_\nu(z)$ is a Bessel function and $K_\nu(z)$ a modified Bessel function. 

Eqs.\   (\ref{possoln})   and   (\ref{negsoln})  contain   the   three
undetermined functions $a(k)$, $b(k)$ and  $c(k)$. The last two can be
expressed  in   terms  of  $a(k)$  by   imposing  suitable  continuity
conditions at $y=0$. From the  BFPE, Eq.\ (\ref{BFPE}), it follow that
both $P(x,y)$ and $\partial_yP(x,y)$ are continuous at $y=0$. Imposing
these conditions gives the relations
\begin{eqnarray}
\label{continuity1}
b(k) & = & K_b\,k^{(\beta_+-\beta_-)/2}\,a(k),  \\
c(k) & = & K_c\,k^{(\beta_--\beta_+)/2}\,a(k),
\label{continuity2}
\end{eqnarray}
where
\begin{eqnarray}
K_b & = & \frac{\pi}{2\sin(\beta_+\pi)}\frac{\Gamma(1+\beta_-)}
{\Gamma(1+\beta_+)}\frac{\left(\beta_+\sqrt{v_+}\right)^{\beta_+}}
{\left(\beta_-\sqrt{v_-}\right)^{\beta_-}}, \\
K_c & = & \frac{1}{2}\,\Gamma(\beta_+)\Gamma(1-\beta_-)
\frac{\left(\beta_-\sqrt{v_-}\right)^{\beta_-}}
{\left(\beta_+\sqrt{v_+}\right)^{\beta_+}}.
\end{eqnarray}

Finally, $a(k)$ is determined by the boundary condition (\ref{bc3}). 
This gives
\begin{eqnarray}
\label{9}
1 & = & \sqrt{-y}\int_0^\infty dk\,\left\{b(k)J_{\beta_-}
\left(2\beta_-\sqrt{kv_-}(-y)^{1/2\beta_-}\right)\right. \nonumber \\
&& + \left. c(k)J_{-\beta_-}\left(2\beta_-\sqrt{kv_-}
(-y)^{1/2\beta_-}\right)\right\},\ \ \ y<0.
\label{bc3app}
\end{eqnarray}
In  principle,   Eqs.\  (\ref{continuity1}),  (\ref{continuity2})  and
(\ref{bc3app}) completely  determine the functions  $a(k)$, $b(k)$ and
$c(k)$.  In  practice, however, we  have been unable to  extract these
functions  explicitly.   We  can,  however,  determine  the  small-$k$
forms. These are sufficient to demonstrate our main claims.

To  determine  the small-$k$  behavior  of  $a(k)$  we consider  Eq.\
(\ref{bc3app}) in  the limit  $y \to -\infty$,  since the  integral is
dominated by small values of $k$ in that limit. We make the ansatz
\begin{equation}
a(k) \sim  A k^a, \ \ \ k \to 0\ .
\label{a}
\end{equation}
Then the small-$k$ forms of $b(k)$ and $c(k)$ are fixed by equations 
(\ref{continuity1}) and (\ref{continuity2}). Since $\beta_+ < \beta_-$, 
$b(k) \gg c(k)$ for $k \to 0$, so the first term, involving $b(k)$, in 
Eq.\ (\ref{bc3app}) dominates for large negative $y$, and the term 
involving $c(k)$ is negligible. In this limit, therefore, Eq.\ (\ref{bc3app})
reads
\begin{eqnarray}
1 & = & AK_b \lim_{y \to -\infty} \sqrt{-y}\int_0^\infty dk\, 
k^{a+(\beta_+ - \beta_-)/2} \nonumber \\
 && \hspace{2cm} \times J_{\beta_-}
\left(2\beta_-\sqrt{kv_-}(-y)^{1/2\beta_-}\right).
\end{eqnarray} 
Evaluating the integral fixes the values of the exponent $a$ and the 
amplitude $A$ in Eq.\ (\ref{a}):
\begin{eqnarray}
a & = & \beta_- - \frac{\beta_+}{2} - 1, \\
A & = & \frac{(\beta_-\sqrt{v_-})^{\beta_-}}{K_b\Gamma(\beta_-)}. 
\end{eqnarray}

Inserting the small-$k$ results for $a(k)$, $b(k)$ and $c(k)$ in Eqs.\ 
(\ref{possoln}) and (\ref{negsoln}) determines $P(x,y)$ in the entire 
regime where $x$ or $y$ (or both) are large. Especially simple is the 
limit of large $x$ at $y=0$, where one obtains
\begin{equation}
P(x,0) \sim \frac{\Gamma(\beta_+)\Gamma(\beta_- - \beta_+)}
{2K_b\Gamma(\beta_-)}\, \frac{(\beta_-\sqrt{v_-})^{\beta_-}}
{(\beta_+\sqrt{v_+})^{\beta_+}}\,\frac{1}{x^{\beta_- - \beta_+}},
\end{equation}
for $x \to \infty$.  This result demonstrates that there is a non-zero
infinite-time survival probability  provided $\beta_- > \beta_+$,  
i.e.\ $\alpha_+ > \alpha_-$. The spatial decay exponent, $\beta_- -  
\beta_+$, should have (we feel) a simple physical  derivation, but so far 
it has eluded us.

\section{Periodic flow fields}
In the remaining part of this  paper we consider the case where $f(y)$
is a periodic function of $y$, with period $2b$. In particular, we 
consider the case
\begin{equation}
f(y) = \cases{v,\ \ \ \ \ \ 2nb < y< (2n+1)b, \cr -v,\ \ \ \ 
(2n-1)b < y < 2nb. \cr}
\end{equation}
for all  integers $n$. This function $f(y)$ is  odd, but we  shall show
that the survival probability decays asymptotically as $t^{-1/2}$. 

\begin{figure}
\narrowtext\centerline{\epsfxsize\columnwidth \epsfbox{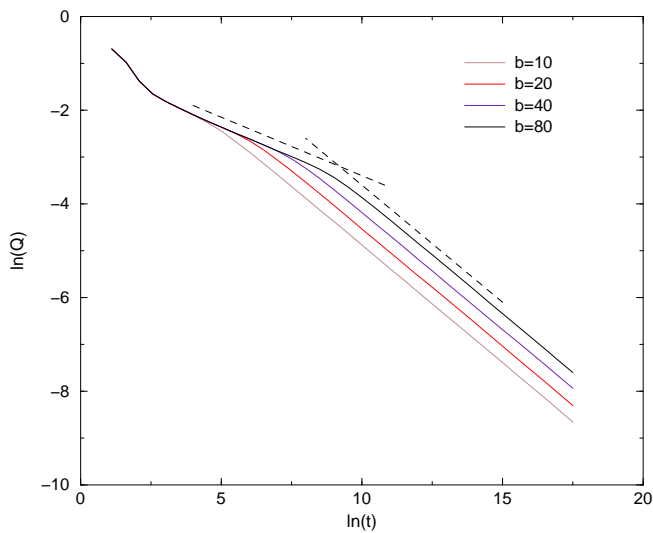}}
\caption{Double  logarithmic  plot  of  the  survival  probability  $Q$
against time $t$ for a range  of band widths $b$. The data show, after
a short-time transient, an initial $t^{-1/4}$ decay crossing over to a
$t^{-1/2}$ decay  at later times.  The dashed lines with slopes $-1/4$
and $-1/2$ are guides to the eye.}
\label{fig1}
\end{figure}

We  first present,  in Figure  \ref{fig1}, the  result of  a numerical
simulation for different values of  the width, $b$, of the alternating
bands. In each case the particle is initially located at $x=0$, on the
interface between two bands.  To avoid immediate absorption, the first
step is taken into the  band with positive velocity.  For early times,
$t  \ll b^2/D$, the  particle explores  the two  bands either  side of
the initial position and the   decay  exponent  $1/4$, appropriate  to
infinitely wide bands  is obtained.  At time $t  \sim b^2/D$, however,
where  the particle  starts  to  explore many  bands, there is a clear
crossover  from   the  initial  $t^{-1/4}$  decay   to  an  asymptotic
$t^{-1/2}$ decay.

To understand these results,  we  first  recall  the use of the Sparre
Anderson theorem \cite{SA} to demonstrate  that $\theta = 1/4$ for odd
functions $f(y)$.  First suppose  that $f(y)>0$ for $y>0$ and $f(y)<0$
for $y<0$.  We focus on crossings of the $x$-axis and regard the steps
between crossings as the  elementary steps of a one-dimensional random
walk on the $x$-axis. The  steps are clearly (i) independent, and (ii)
drawn  from  a  symmetric  distribution. The  Sparre Anderson  theorem
states  that the  probability that  the  process has  not crossed  the
absorbing  boundary at  $x=0$  after  $N$ steps  decays  as $Q_N  \sim
N^{-1/2}$ for  large $N$.  Since the  number of crossings  in time $t$
scales as $N \sim t^{1/2}$, this implies $Q(t) \sim t^{-1/4}$.

Let  us examine this  argument more  carefully. It's  validity clearly
requires that a crossing of  the absorbing boundary at $x=0$ can occur
at most once between two  consecutive crossings of the $x$-axis.  This
will always  be the case if $f(y)$  takes only one sign  for $y>0$ and
the other sign for $y<0$.  In the periodic case, however, $f(y)$ takes
both  signs for  $y>0$  and both  signs  for $y<0$.   The process  can
therefore cross the absorbing boundary, from $x>0$ to $x<0$, and cross
back  again  {\em without}  crossing  the  $x$-axis.  Such  absorption
process  will  be missed  in  a  description  which focuses  only  on
crossings  of  the  $x$-axis.    Such  a  description  will  typically
underestimate  the  value of  $\theta$  (it  gives  a lower  bound  on
$\theta$).

A better way of looking at this problem is as follows. The probability
$Q(x,y,t)$ that  the particle survives  until time $t$, given  that it
started at  $(x,y)$, is periodic  with period $2b$. In particular, the
partial derivative $\partial_yQ$ vanishes by symmetry at the center of
any band.   It therefore suffices to  solve the problem  in the region
$-b/2 \le y \le b/2$, $x  \ge 0$, which defines a semi-infinite strip,
with boundary conditions $Q(0,y,t)=0$, $\partial_yQ(x,\pm b/2,t) = 0$.
The flow  velocity is $v$ in the  upper part of the  strip ($y>0$) and
$-v$  in the  lower part.  This strip  problem has  been  discussed by
Redner and Krapivsky \cite{RK}, who argue as follows. The typical time
between crossings of the center line, $y=0$, is $\tau \sim b^2/D$. The
typical distance travelled,  in the $x$-direction, in this  time is $l
\sim  vb^2/D$. The two cases $x \gg l$ and $x \ll l$, where $x$ is the 
initial displacement of the particle, have to be analysed separately. 
The following discussion is based on reference \cite{RK}.

\subsection{The regime  $x \gg l$} 
The crossings  of  the  $x$-axis define  a symmetric  one-dimensional  
random  walk  with  step  length $\sim l$  and time step $\sim \tau$.  
The effective diffusion constant for motion  in the $x$-direction is,   
therefore,    $D_\parallel \sim l^2/\tau \sim v^2b^2/D$. The survival  
probability for this random walk  is given by the well-known result 
\cite{Guide}
\begin{equation}
Q(x,y,t) \sim \frac{x}{\sqrt{D_\parallel t}} \sim 
\frac{x}{bv}\sqrt{\frac{D}{t}}.   
\label{1d}
\end{equation}
Note that the $y$-dependence is not determined by this argument. It is
intuitively clear, however,  that the result is insensitive  to $y$ at
large $x$. 

Comparing  this  result  with  Figure  \ref{fig1}, we  note  that  the
amplitude  of  the  $t^{-1/2}$   behaviour  in  the  late-time  regime
increases  with  increasing $b$  in  Figure  \ref{fig1}, whereas  Eq.\
(\ref{1d}) predicts  a decreasing  amplitude with increasing  $b$. The
data, however, were obtained with  the initial $x$ equal to zero (with
$x=v$ after one  time step).  To make meaningful  comparisons with the
data, therefore, we need to analyse the opposite regime $x \ll l$.

\subsection{The regime $x \ll l$}
\begin{figure}
\narrowtext\centerline{\epsfxsize\columnwidth \epsfbox{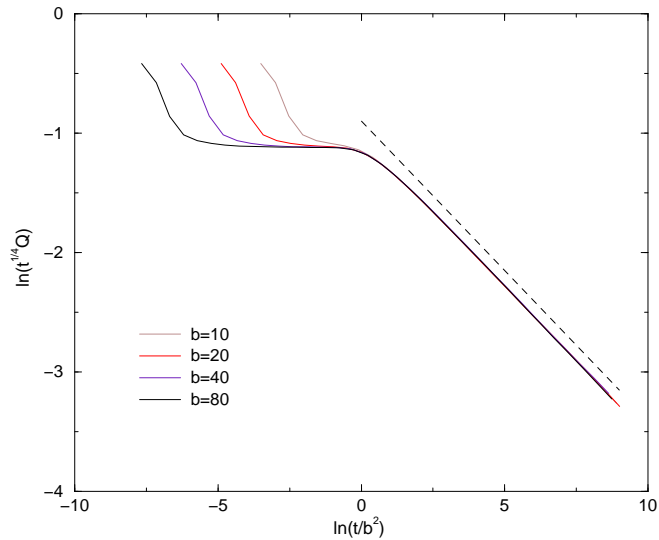}}
\caption{Double  logarithmic plot  of  the  survival probability 
$Q$ against time $t$ plotted in the scaling form $t^{1/4}Q$ against 
$t/b^2$. The dashed line with slopes $-1/4$ is a guide to the eye.}
\label{fig2}
\end{figure}

For $x \ll l$, one can argue as follows. At short times, $t \ll \tau$,
the  behaviour  will  be  the  same  as  for strips of infinite width, 
$Q(t)\sim (x/vt)^{1/4}$ \cite{RK}. At later times, this result will be 
modified by a function of $t/\tau$, $Q(t)=(x/vt)^{1/4}f(Dt/b^2)$.  For 
$t \gg\tau=b^2/D$, there should be a crossover to the $t^{-1/2}$ decay
calculated above, since  if the particle survives for  time $\tau$, it
will  typically reach  a  distance $x  \sim  v\tau =  l$. The  scaling
function $f(z)$ therefore behaves as $z^{-1/2}$ for large $z$, giving
\begin{equation}
Q(x,0,t) \sim \left(\frac{xb^2}{vD}\right)^{1/4} \frac{1}{t^{1/2}}, \ \ \ 
t \gg b^2/D\ .
\label{1d'}
\end{equation}
The two  results Eqs.\ (\ref{1d}) and (\ref{1d'})  match, as required,
at $x \sim l=vb^2/D$, where  both reduce to $Q \sim b/\sqrt{Dt}$.  The
scaling  form  $Q(t)  \sim  t^{-1/4}f(Dt/b^2)$  is  tested  in  Figure
\ref{fig2}, where  $t^{1/4}Q$ is plotted  against $t/b^2$ on  a double
logarithmic  plot.  After the  initial  transients,  there  is a  good
collapse of the  data onto a scaling curve, in  which an initial slope
of  zero gives  way,  at $t \simeq  b^2$, to a  final slope of $-1/4$,
corresponding  to  the   regimes  $Q(t)\sim  t^{-1/4}$  and  $Q(t)\sim
t^{-1/2}$ respectively.

\begin{figure}
\narrowtext\centerline{\epsfxsize\columnwidth \epsfbox{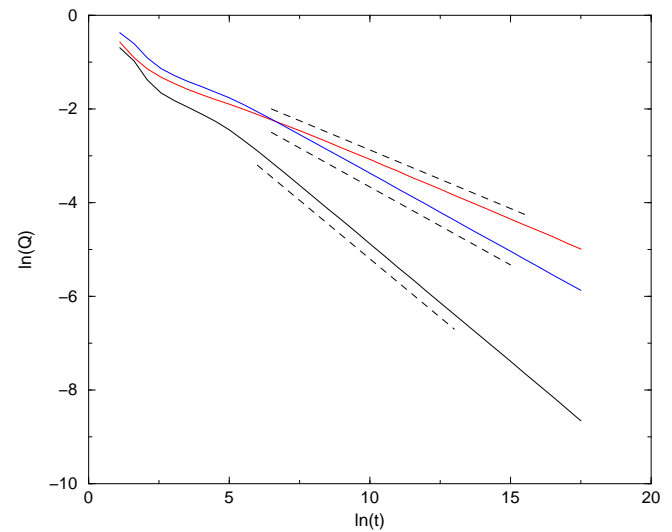}}
\caption{Double  logarithmic  plot  of  the  survival probability  $Q$
against time $t$ for the band model, for different types  of absorbing 
boundary. The dashed lines with slope $-1/4$,  $-1/3$ and  $-1/2$  are 
guides to the eye. In order of increasing steepness, the corresponding 
data refer to absorbing boundaries given by (i) the negative $y$-axis,  
(ii) the  negative  $y$-axis and the  negative  $x$-axis and (iii) the 
whole $y$-axis.  Least squares fits to the data give slopes -0.254, 
-0.333 and -0.501 respectively, consistent with the predicted values. 
}
\label{fig3}
\end{figure}

We  can interpret the asymptotic result $Q \sim t^{-1/2}$ as follows.  
If we coarse grain  in the $y$-direction,
the average deterministic drift is zero, and the nominally subdominant
diffusive  motion  in  the  $x$-direction  \cite{RK},  with  diffusion
constant  $D_\parallel$,  plays  an  important  role.   The  effective
coarse-grained  dynamics  corresponds  to anisotropic  two-dimensional
diffusion. If the absorbing  boundary is the whole $y$-axis, diffusion
in the  $y$-direction is irrelevant  (except for its role  in inducing
diffusion  in  the  $x$-direction)  and the  standard  one-dimensional
result  for the  survival probability,  Eq.\ (\ref{1d}),  is obtained.
One can now discuss, however, other types of absorbing boundary. First
we  consider  absorption on  the  half-line  $x=0$, $y<0$.   Numerical
simulations, presented in Figure  \ref{fig3}, suggest   $\theta = 1/4$
for this case, while for absorption on the two half lines $x=0$, $y<0$
and $y=0$, $x<0$, which define an absorbing wedge of angle $\pi/2$, 
the data suggest (for $x>0$ initially) $\theta = 1/3$.

These  seemingly mysterious  exponents are  readily understood  in the
context  of  the   underlying  anisotropic  two-dimensional  diffusion
process. For {\em isotropic} diffusion inside a wedge of opening angle
$\phi$,  with absorbing  boundaries on  the  edges of  the wedge,  the
survival  probability is known  to decay  as a  power law,  $Q(t) \sim
t^{-\theta}$,  with $\theta =  \pi/2\phi$ \cite{Fisher-Gelfand,Guide}.
The  absorbing  boundary on the half-line $x=0$,  $y<0$ corresponds to
opening angle $\phi=2\pi$, giving  $\theta = 1/4$, while the absorbing
boundary  on  the  two   half-lines  $x=0$,  $y<0$  and  $y=0$,  $x<0$
corresponds  to $\phi=3\pi/2$,  giving $\theta  = 1/3$.   Although the
effective diffusion process is here anisotropic, it can be transformed
into an isotropic process by rescaling the $x$ or $y$ variable.  Since
this  rescaling does  not change  the  geometry of  the two  absorbing
boundaries considered, the results $\theta=1/4$ and $1/3$ respectively
are not changed.

It is straightforward to apply  the same general reasoning to the band
model where the positive- and negative-flow bands are inequivalent.   
Consider a model where bands of width $b_+$ and $f(y) = v_+$ alternate 
with bands of width $b_-$ and $f(y)=-v_-$. For the special cases where 
$b_+v_+=b_-v_-$ there will be no macroscopic flow, and a coarse-grained
description will lead to an anisotropic diffusion model as before, with
the same values of $\theta$, e.g.\ $\theta=1/2$ if  the whole $y$-axis
is absorbing (indeed, $\theta=1/2$ will hold for any periodic $f(y)$ 
with zero mean). If $b_+v_+ -  b_-v_- >0$, the coarse-grained model has
a mean  flow away from  the absorbing boundary, and  the infinite-time
survival probability is non-zero, while  for $b_+v_+ - b_-v_- < 0$ the
mean flow is towards the  boundary and the survival probability decays
exponentially  with time. In  the following  subsection we  provide an
exact solution for  the case $b_+v_+ = b_-v_-$, in  the limit $b_- \to
0$, $v_- \to \infty$, with $u = b_-v_-$ held fixed.

\subsection{A solvable model}
The model we study is defined by a periodic function $f(y)$, with 
period $b$, given by
\begin{equation}
f(y) = v - u\,\delta(y), \ \ \ \ -b/2 \le y \le b/2,
\end{equation}
with  $f(y+b)=f(y)$  for all  $y$.   The  marginal  case of  interest,
corresponding to no net flow, is given by $u=bv$. It is convenient, in
the  first instance,  to compute  the killing  probability $P(x,y,t)$.
The periodicity implies that we  need only consider the regime $-b/2 \le
y  \le  b/2$, with  the  boundary  conditions $\partial_yP(x,b/2,t)  =
\partial_yP(x,-b/2,t)=0$. The function $P$ obeys the BFPE 
\begin{equation}
\partial_tP = D\partial_{yy}P + [v-u\delta(y)]\partial_xP.
\label{toy}
\end{equation}
Taking the Laplace transform with respect to time, and exploiting the 
initial condition $P(x,y,0)=1$ for all $x>0$, gives
\begin{equation}
D \partial_{yy}\tilde{P}+[v-u\delta(y)]\partial_x\tilde{P} - s\tilde{P}=0,
\label{laplace}
\end{equation}
where $\tilde{P}(x,y,s)=\int_0^\infty dt\,\exp(-st)P(x,y,t)$. The general 
solution satisfying the given boundary conditions at $y=\pm b/2$ is 
(setting $D=1$ for convenience) 
\begin{eqnarray}
\tilde{P}(x,y,s) & = & \int_0^\infty dk\,a(k,s)e^{-kx} \nonumber \\
&& \hspace{1cm} \times \cosh\left[\sqrt{kv+s}\,
\left(\frac{b}{2}-|y|\right)\right],
\label{ptilde}
\end{eqnarray}
for $y \ne 0$. Imposing the discontinuity in $\partial_y\tilde{P}$ at 
$y=0$ implied by Eq.\ (\ref{laplace}), $\partial_y\tilde{P}(x,0+,s) - 
\partial_y\tilde{P}(x,0-,s) = u\partial_x\tilde{P}(x,0,s)$ for all $x>0$ 
and all $s$, leads to the condition
\begin{equation}
\tanh \gamma = \frac{ub}{\gamma v}\left(\frac{\gamma^2}{b^2} 
- \frac{s}{4}\right),
\label{gammaeqn}
\end{equation}
where
\begin{equation}
\gamma = \frac{b}{2}\sqrt{kv+s}.
\label{gamma}
\end{equation}
Eq.\ (\ref{gammaeqn}) shows that only a single value of $k$ is possible 
for each $s$. Specialising to the marginal case $u=bv$, this equation 
simplifies to 
\begin{equation}
\tanh \gamma = \gamma - \frac{b^2s}{4\gamma}\ .
\label{gamma1}
\end{equation}
The large-time behavior is governed by the small-$s$ solution of Eq.\ 
(\ref{gamma1}): $\gamma \simeq (3b^2s/4)^{1/4}$. Inserting this into 
Eq.\ (\ref{gamma}) gives, for small $s$, $k \simeq 2\sqrt{3s}/bv$ and 
$\sqrt{kv+s} \simeq (3s)^{1/4}\sqrt{2/b}$. Putting these results into 
Eq.\ (\ref{ptilde}) gives
\begin{eqnarray}
\tilde{P}(x,y,s) & = & a(s)\,\exp\left(-\frac{2\sqrt{3s}}{bv}\,x\right) 
\nonumber \\
&& \hspace{1cm}\times \cosh\left(\frac{(3s)^{1/4}}{\sqrt{b/2}}
\left(\frac{b}{2}-|y|\right)\right),
\end{eqnarray}
for $s \to 0$.

The amplitude $a(s)$ is  fixed by the boundary condition, $P(0,0,t)=1$
for  all  $t$,  which  gives  $\tilde{P}(0,0,s) =  1/s$  and  $a(s)  =
\{s\cosh[(3s)^{1/4}\sqrt{b/2}]\}^{-1}$   for    small   $s$.    Since
$Q(x,y,t)  =  1  -  P(x,y,t)$,  we  have  $\tilde{Q}(x,y,s)  =  1/s  -
\tilde{P}(x,y,s)$. The small-$s$ result for $\tilde{P}$ determines the
large-$t$ form of $Q(x,y,t)$ as
\begin{equation}
Q(x,y,t) = \frac{2\sqrt{3}}{b\sqrt{\pi t}}\,\left(\frac{x}{v} 
+ \frac{1}{2} \,|y|(b - |y|)\right),
\label{exact}
\end{equation} 
for $|y|  \le b/2$. This result  confirms, for this  special case, the
result  $Q(t) \sim  t^{-1/2}$ obtained  above using  general heuristic
arguments for  the case where there  is no net flow.  Note that Eq.\ 
(\ref{exact}) has the same form as Eq.\ (\ref{1d}), and that this form
holds for  all values of $x/l$,  where $l=v b^2/D$, i.e.\  there is no
analogue, in  this model, of Eq.\  (\ref{1d'}) for $x \gg  l$. This is
because in the  present model there is no regime  in which the systems
resembles a two-band model.

\section{Discussion and Summary}
In this  paper we have  obtained further results  on the first-passage 
properties of a particle  diffusing in the  $y$-direction and  subject 
to  a deterministic drift velocity, $f(y)$, in the $x$-direction, with 
an absorbing boundary at $x=0$. 

In the  first part  of the  paper we considered  the case  where $f(y)
\propto y^{\alpha_+}$ for $y>0$ and $f(y) \propto -|y|^{\alpha_-}$ for
$y<0$.  We confirmed the  conjecture proposed  in \cite{GB04}  that the
larger  power is `strongly  dominating' (in  contrast  to the  `weakly
dominating'  behaviour  obtained  when  the exponents  $\alpha_+$  and
$\alpha_-$ are equal but the amplitudes are different \cite{GB04}). We
showed  explicitly  that  for   $\alpha_+  >  \alpha_-$  the  survival
probability approaches  a non-zero limiting value, and  we obtained an
expression for this value.

In the  second part of the  paper we considered the  case where $f(y)$
has  a banded  periodic  form,  with alternating  bands  of equal  and
opposite drift  velocity. We presented numerical data,  supported by a
heuristic argument,  to show that the  asymptotic persistence exponent
is  $\theta = 1/2$  in this  case, and  obtained a  generalisation for
different  forms of  absorbing  boundary. We  argued  that the  result
generalises to any model where the coarse-grained flow field vanishes,
and exemplified this result through a soluble model.

PG's work was supported by EPSRC.

\end{multicols}


\begin{references}     

\bibitem{GB04} 
A. J. Bray and P. Gonos, J.\ Phys.\ A {\bf 37}, L361 (2004). 

\bibitem{RK} 
S. Redner and P. L. Krapivsky, J. Stat.\ Phys.\ {\bf 82}, 999 (1996).

\bibitem{Burkhardt} 
T. W. Burkhardt, J. Phys.\ A {\bf 33}, L429 (2000). 

\bibitem{Guide} 
S. Redner, {\em A Guide to First-Passage Processes} (CUP, Cambridge, 2001). 

\bibitem{majumdar_review} 
S. N. Majumdar, Curr.\ Sci.\ India {\bf 77}, 370 (1999). 

\bibitem{SA} 
E. Sparre Andersen, Math.\ Scand.\ {\bf 1}, 263 (1953); 
{\em ibid.} {\bf 2}, 195 (1953). 

\bibitem{Fisher-Gelfand}
M. E. Fisher and M. P. Gelfand, J.\ Stat.\ Phys.\ {\bf 53}, 175 (1988). 


\end{references}
\end{document}